\documentclass[sigconf, screen]{acmart}

\AtBeginDocument{%
  }

\copyrightyear{2026}
\acmYear{2026}
\setcopyright{cc}
\setcctype{by-nc-sa}

\acmConference[AutomationXP26 Workshop at CHI '26]{AutomationXP26 @ CHI '26}{April 14, 2026}{Barcelona, Spain}
\acmBooktitle{AutomationXP26 Workshop at CHI '26}

\acmDOI{}                          

\acmISBN{}                         
\acmPrice{}                        

\usepackage{enumitem}

\begin{document}

\title{From Human Negotiation to Agent Negotiation: Personal Mobility Agents in Automated Traffic}
\thanks{Position paper presented at AutomationXP26: Agentic Automation Experiences — Rethinking the Interaction of Humans and AI Agents at CHI~’26}

\author{Pascal Jansen}
\email{pascal.jansen@uni-ulm.de}
\orcid{0000-0002-9335-5462}
\affiliation{%
  \institution{Institute of Media Informatics, Ulm University}
 ~\city{Ulm}
  \country{Germany}
}

\renewcommand{\shortauthors}{Pascal Jansen}

\begin{abstract}
Conflicts between user preferences and automated system behavior already shape the experience of automated mobility. For example, a passenger may prefer assertive driving, yet the vehicle slows down early to follow a conservative policy or yield to other actors. Similar conflicts arise at merges, crossings, or right-of-way situations, where users must accept opaque decisions or attempt to negotiate through interfaces not designed for continuous, multi-actor relationships.
This position paper argues that such approaches do not scale as mobility becomes more heterogeneous and automated. Instead, it proposes \textit{personal mobility agents} that act as proxies for users, encode preferences such as comfort and safety margins, and negotiate traffic behavior with other agents under shared safety rules. The central idea is a shift from moment-to-moment user negotiation interfaces to delegation and oversight interfaces, in which proxy agents manage real-time conflicts while users can shape high-level policies and preferences.
\end{abstract}



\maketitle


\section{Introduction}

Mobility is a foundational socio-technical infrastructure because it enables access to work, education, healthcare, and social participation, while failures of mobility systems directly translate into safety risks~\cite{world2024global}. At its core, everyday traffic is a continuous negotiation among multiple actors (e.g., drivers and cyclists) over shared space, including who proceeds first, what constitutes an acceptable gap, and how a driving maneuver is executed.

Automated mobility changes this negotiation by transferring parts of decision-making from humans to automated systems. Consequently, prior work has focused on how automated systems communicate with surrounding road users, for example, via external human--machine interfaces (eHMIs) that signal intent to pedestrians~\cite{deWinter2022eHMINecessity}. Across these strands, \textit{trust calibration} (see \citet{LeeSee2004Trust}) remains a central design objective as users must neither over-rely on automation nor avoid it, and adoption at scale depends on perceived safety, reliability, and preference alignment~\cite{Wischnewski2023TrustCalibration, deWinter2022eHMINecessity}.
Beyond signaling intent, recent work has also explored negotiation interfaces in automated driving, for example, in passenger–vehicle conversational negotiation scenarios~\cite{Stampf2024LLMNegotiation}. These approaches move beyond static communication by allowing users to influence or discuss momentary decisions.

However, many interactions assume a one-human--one-system negotiation relationship, for example, in pedestrian--vehicle interaction at crossings~\cite{deWinter2022eHMINecessity} or in passenger--vehicle personalization such as driving-style adaptation~\cite{Vasile2023DrivingStyle}. In contrast, prior work in multi-agent negotiation explores the design of protocols and negotiation strategies among automated actors in traffic settings, providing a formal basis for shared decision-making and maneuver coordination~\cite{belloni2015dealing}. However, these approaches do not model persistent, user-representing preference proxies or delegation interfaces across situations.

In practice, automated traffic is a multi-actor environment involving drivers, passengers, pedestrians, automated vehicles, micromobility users (e.g., e-scooter riders), and intelligent infrastructure. Actors differ in abilities, preferences, and goals, such as age-related differences in gap acceptance~\cite{Lobjois2007StreetCrossing}, cultural differences in risk tolerance~\cite{HellJapanGermanyReview}, and preferences for automated driving style~\cite{Vasile2023DrivingStyle}. Hence, traffic is a multi-objective negotiation under time pressure and partial observability, where each actor optimizes different objectives with incomplete information about others~\cite{deWinter2022eHMINecessity}. These conditions make conflicts between user intent and system constraints likely and can reduce automation engagement, trust, and acceptance~\cite{stampf2024conflicthandling}.

To exemplify such conditions, consider an urban traffic scenario in the near future:
\begin{quote}
A passenger sits in an automated ride-sharing vehicle approaching a crossing. A cyclist approaches from the side, while a delivery robot moves along the sidewalk, and a pedestrian prepares to cross. The vehicle slows down earlier than the passenger expects, apparently yielding to the pedestrian. The cyclist, however, accelerates and crosses first, while the delivery robot pauses due to its own internal policy. From the passenger’s perspective, the situation appears inconsistent and difficult to interpret: it is unclear why the vehicle behaved conservatively, whether this reflects their own preferences, the cyclist’s behavior, or hidden infrastructure data. The user has no practical way to express a preference, such as “be more assertive in light traffic but always keep large safety gaps near pedestrians,” nor to understand how this preference would be negotiated with other actors.
\end{quote}
In such scenarios, existing interaction approaches (e.g.,~\cite{Stampf2024LLMNegotiation}) would require either additional human-facing explanations, explicit negotiation dialogues, or post-hoc justifications. However, these typically assume that users have appropriate \textit{situation awareness} (see \citet{Endsley1995SA}) so that they can meaningfully track and evaluate each interaction. In increasingly heterogeneous environments, such assumptions may not hold at scale. This could lead to disengagement, more overrides, generalized distrust, and ultimately reduced perceived control, trust calibration, and acceptance of automated mobility.

This position paper, therefore, argues that scalable mobility interaction will require shifting from direct user-to-user negotiation to proxy-mediated negotiation. It therefore proposes a different interaction model in which each road user is represented by a \textbf{personal mobility agent} that acts as a proxy between the user and the traffic situation. The agent encodes the user’s preferences, constraints, and delegation policies, and negotiates with other actors on the user’s behalf. For example, a pedestrian’s agent may request larger safety gaps, while a vehicle’s agent may negotiate a smoother but slightly slower merge to preserve passenger comfort. Instead of requiring the user to participate in each momentary negotiation, the agents resolve these interactions among themselves and communicate only high-level outcomes to their users, if desired, such as “yielding early to maintain your preferred safety margin.”

In this view, interaction shifts from direct control of individual maneuvers toward the specification of high-level preferences and delegation policies. The design problem then becomes ensuring that proxy-based negotiation mechanisms remain safe, understandable, and acceptable to all involved actors.

\section{Personal Mobility Agents in Automated Traffic Negotiations}

This section outlines the possible technical foundations of personal mobility agents, describes potential mechanisms for their personalization, and exemplifies how they could operate in future automated traffic scenarios.

\subsection{Technological Foundations}

From a technological perspective, personal mobility agents could build on several current developments. Automated vehicles and driver-assistance systems can already learn individual driving preferences, such as comfort levels or acceleration styles, from user input and observed behavior~\cite{liao2025drivingstyle}. Human-in-the-loop optimization approaches further show that systems can learn such preferences efficiently through short interactive sessions. For example, Bayesian optimization has been used to adapt eHMIs and in-vehicle interfaces to individual users~\cite{jansen2025opticarvis, colley2025improving}. In parallel, recent progress in agentic AI, large language models, and vision–language models suggests that software agents can act on behalf of users, maintain persistent preference models, and interact with other agents or services~\cite{Park2023GenerativeAgents}. Many of these systems can be personalized using reinforcement learning from human feedback, where the system learns preferred behavior from human ratings rather than fixed rules~\cite{OpenAI2025Operator}. A personal mobility agent can be understood as a specialization of such personal agents for traffic contexts.

\subsection{Core Functional Components}

A personal mobility agent may perform three coupled functions:

    \textbf{Preference modeling}
    The agent maintains a \textit{preference model}, akin to a digital twin of the user, that captures aspects such as acceptable safety gaps, preferred acceleration levels, comfort limits, willingness to yield, and tolerance for delays or detours. This model, however, might remain constrained by safety and legal requirements (e.g., \cite{GermanyAutonomousDrivingAct2021}).

    \textbf{Traffic negotiation}
    During traffic events such as approaching a crossing, merging into a lane, following a slower vehicle, or entering a roundabout, the agent evaluates feasible maneuvers under safety constraints and negotiates a coordination outcome with other automated actors. For example, when two vehicles approach a merge, each agent may propose a maneuver that reflects its user’s preferences, such as “yield early for comfort” or “merge efficiently if the gap is safe.” The resulting maneuver is negotiated under shared safety rules, while the user remains at the high-level preference and policy control level.

    \textbf{Interface management}
    The agent manages the \textit{human-facing interface} by communicating outcomes. For example, after a merge, it may report: “merged conservatively due to comfort policy,” or after approaching a crossing: “yielded early to maintain your preferred safety margin.” This follows the principle of calibrated trust, where users receive intelligible feedback about system behavior~\cite{LeeSee2004Trust, Wischnewski2023TrustCalibration}. In addition, the interface would support \textit{selective disclosure and contestability}, as it summarizes relevant outcomes, reveals trade-offs when behavior deviates from expectations, and allows users to inspect and revise decision policies.

\subsection{Lifecycle of a Personal Mobility Agent}

From the user’s perspective, the personal mobility agent might be created and refined in three phases:
\begin{enumerate}[topsep=2pt,itemsep=2pt,parsep=0pt,leftmargin=*]
    \item During setup, the agent initializes the \textit{preference model} using simple onboarding questions (e.g., rating comfort versus efficiency preferences in traffic), demographics, personality tests, and/or a short calibration drive or simulation.
    \item The agent then adapts online during regular use. It observes how the user reacts to different maneuvers, such as whether they accept or override certain behaviors, and occasionally asks for quick feedback.
    \item The agent is updated over longer time scales. It detects changes in user behavior, allows explicit policy edits (e.g., “be more assertive in light traffic”), and periodically recalibrates its preference model. The agent is therefore not a fixed model trained once, but an evolving representation that is refined both offline and online while serving as the user’s proxy.
\end{enumerate}

\section{Design Tensions and Research Directions}

Introducing personal mobility agents that act as proxies for users shifts interaction design away from isolated control interfaces toward the design of \textit{negotiation representations} and \textit{delegation policies}. In this model, the user is no longer directly involved in each traffic interaction. Instead, their proxy negotiates on their behalf. This creates several design tensions that may become relevant in future mobility systems and automation experiences.

\subsection{Designing for Preference Alignment Under Legal and Safety Constraints}

Mobility systems are constrained by strict safety and legal requirements, which may force the proxy to deviate from users' preferences. For example, a user may prefer small gaps when merging, but the proxy must still enforce safety thresholds.

Prior negotiation-focused HMI approaches often attempt to resolve such conflicts through explanation, persuasion, or adaptive dialogue~\cite{stampf2024conflicthandling, Stampf2024LLMNegotiation}. In a proxy-interaction model, however, the design problem shifts from resolving each conflict interactively in the moment to constraining policy space. This raises the question of how the agent should communicate differences between the user’s stated preferences and the behavior it executes, and whether such communication would undermine the intended benefits of relieving users from direct interaction in momentary situations.

\subsection{Preventing Strategic Manipulation While Preserving Personalization}

Research on LLM-mediated negotiation suggests that users appreciate adaptive communication, but may also attempt to probe or outsmart the system~\cite{Stampf2024LLMNegotiation}. This issue may become more critical with persistent agents that accumulate personalization over time and repeatedly act on the user’s behalf.
Research on delegation in human–automation decision contexts suggests that users’ willingness to delegate depends on perceived competence, controllability, and accountability of the delegated system~\cite{spitzer2025human}. Persistent proxy agents, therefore, introduce long-term governance challenges beyond momentary adaptation.

A possible direction is to separate \textit{interaction adaptivity} (how the agent communicates) from \textit{policy adaptivity} (what the agent is allowed to do). This suggests design requirements for manipulation-resistant preference capture, non-negotiable safety rules, and interaction patterns that discourage adversarial bargaining.

\subsection{Scaling Agent Negotiation Without Breaking Mixed-Traffic Interaction}

Personal mobility agents could enable efficient and effective negotiation among automated actors by exchanging preferences and constraints as data points. However, traffic will likely remain a mixed environment in which many actors lack such an agent and behave unpredictably~\cite{deWinter2022eHMINecessity}. A proxy must therefore interact not only with other agents but also with human drivers, cyclists, and pedestrians, whose intentions must be inferred from their behavior.
Prior work on eHMIs has highlighted that not all actors share symmetric communication channels or shared representations~\cite{deWinter2022eHMINecessity}. Proxy-based coordination, therefore, cannot assume universal agent participation and must degrade gracefully to behavior-based inference.

This creates a tension between optimizing agent-agent negotiations and maintaining robust interaction with human actors. In addition, proxy-based negotiation may obscure responsibility when outcomes emerge from multiple interacting agents. A possible research direction is the design of \textit{attribution mechanisms} that explain whether a behavior was driven by safety constraints, the user’s delegation policy, inferred human behavior, or a negotiated compromise with other proxies.
This challenge echoes broader concerns in multi-agent automation about distributed responsibility and attribution when outcomes emerge from interacting autonomous systems rather than a single decision-maker (e.g., see~\cite{belloni2015dealing}).

\subsection{Making Delegation Legible: Transparency, Agency, and Accountability}

Proxy-based interactions change not only system behavior but also the user’s experience with automation. When a proxy negotiates most situations autonomously, users may become less aware of how decisions are made, which trade-offs are applied, or which constraints shape outcomes. In some cases, this reduced involvement may lower cognitive load and might even increase acceptance because the system “just works.” In other cases, however, it may lead to loss of agency, miscalibrated trust, or difficulties in understanding why certain outcomes occurred.

This creates an ethical and experiential tension between \textit{convenient delegation} and \textit{meaningful awareness}. Designers must decide how visible proxy decisions should be. One option is minimal disclosure, where the proxy operates largely in the background and only surfaces important deviations or safety-critical events. Another option is reflective transparency, where the system periodically exposes negotiation summaries or policy trade-offs to support user understanding and long-term trust calibration.

This tension raises several open questions: Do users prefer to remain largely unaware of negotiation details as long as outcomes match their expectations? At what point does reduced awareness undermine perceived agency or accountability? And how should proxy systems balance cognitive load reduction with ethical requirements for transparency, contestability, and informed consent?

\section{Conclusion}

This position paper argues that traffic in the near future will become too heterogeneous and automated for users to meaningfully negotiate each traffic situation themselves. As automated vehicles, micromobility, robots, and connected infrastructure interact in shared spaces, conflicts between actors with different goals and preferences become inevitable, while traditional moment-to-moment human negotiation strategies struggle to scale.
Instead, \textit{personal mobility agents} may act as proxies for users. These agents encode user preferences and constraints and negotiate concrete maneuvers with other actors under shared safety rules.

This shift raises relevant research questions for HCI. Proxy-based interactions may reduce cognitive load and improve acceptance, but they may also change how users perceive agency, responsibility, and trust when most negotiations occur without their direct involvement. Understanding how to design delegation interfaces, how much awareness users need, and how proxy-mediated experiences affect long-term trust and accountability remains an open research agenda for automated mobility.
Similar proxy-mediated interaction models may emerge in domains such as domestic robots, smart buildings, and collaborative AI systems.


\bibliographystyle{ACM-Reference-Format}
\bibliography{mobility-proxy-agents}


\end{document}